\documentclass[aps,prl,twocolumn,floatfix,showpacs,superscriptaddress]{revtex4}
\usepackage{graphicx}
\usepackage{amsmath}
\begin{document}

\title{Single-photon exchange interaction  
in a semiconductor microcavity}
\author{G. Chiappe}
\affiliation{Departamento de F\'{\i}sica Aplicada,
Universidad de Alicante, San Vicente del Raspeig, Alicante 03690, Spain.} 
\affiliation{Departamento de  F\'{\i}sica J.J. Giambiagi, Facultad de
Ciencias Exactas, Universidad de Buenos Aires, Ciudad Universitaria,
1428 Buenos aires, Argentina.}
\author{J. Fern\'andez-Rossier}
\affiliation{Departamento de F\'{\i}sica Aplicada,
Universidad de Alicante, San Vicente del Raspeig, Alicante 03690, Spain.} 
\author{E. Louis}
\affiliation{Departamento de F\'{\i}sica Aplicada,
Universidad de Alicante, San Vicente del Raspeig, Alicante 03690, Spain.} 
\affiliation{Instituto Universitario de Materiales and 
Unidad Asociada del CSIC, 
Universidad de Alicante, San Vicente del Raspeig, Alicante 03690, Spain.} 
\author{E.V. Anda}
\affiliation{Departamento de  F\'{\i}sica, Pontificia Universidade
Cat\'olica do Rio de Janeiro (PUC-Rio), 22452-970, Caixa Postal:
38071 Rio de Janeiro, Brazil.}
\date{\today}

\begin{abstract}
We consider the effective coupling of localized spins  in a semiconductor
quantum dot embedded in a  microcavity. The lowest cavity mode and the quantum
dot exciton are coupled and close in energy, forming a polariton. The fermions
forming the exciton   interact 
with localized spins via  exchange. Exact diagonalization of a
Hamiltonian in which photons, spins and excitons are treated quantum
mechanically  shows that  {\it a single  polariton} induces a
sizable indirect exchange  interaction between otherwise independent spins. The origin, symmetry
properties and the intensity of that  interaction depend both on the dot-cavity
coupling and detuning.  In the case of a (Cd,Mn)Te quantum dot, Mn-Mn
ferromagnetic coupling  mediated by a single photon survives above 1 K whereas
the exciton mediated coupling survives at 15 K.  

\end{abstract}
\maketitle

Control of exchange interactions in solid state environments has  become a
strategic target in the development of   both  spintronics and   quantum
computing  \cite{LD98,SI99,Ka98,WA01,AL02}.  Artificial control of {\em direct}
exchange interactions, which occur  at  length scales of one  lattice spacing,
is hardly  possible with current day technologies. In contrast, there is a
number of proposals to control {\em indirect} exchange interactions (IEI) of
distant  spins sitting several nanometers away
\cite{LD98,SI99,Ka98,PC02,Merlin02,Ca03,FB04,FP04}, taking advantage of the
optical and electrical manipulation of the intermediate fermions afforded  in
semiconducting hosts. The local spins could be provided by the nuclei
\cite{Ka98}, by electrons bound to donors \cite{PC02,Merlin02}, or $d$
electrons of magnetic impurities \cite{FB04,FP04}. Artificial control of the
IEI has been observed experimentally giving rise to a variety of phenomena,
like the reversible modification of the Curie temperature and coercive fields 
in (III,Mn)V \cite{Nature} and (II,Mn,N)VI semiconductors \cite{BK02}, the
induction magnetic order in otherwise paramagnetic (II,Mn)VI semiconductor
quantum dots \cite{dot-Smith} and the enganglement of donor spins in (II,Mn)VI
quantum wells \cite{Merlin02}.  Such a control is also a must in the
implementation of  quantum computation using localized spins in solids, since 2
qbit operations require exchange interactions between distant spin pairs
\cite{Ba95,Di00}.

Laser driven IEI in semiconductors is particularly promising because it affords
control in the time domain and it can be tuned by changing the laser frequency,
intensity and polarization.  Above gap excitation \cite{Merlin02,BK02,dot-Smith}
creates real carriers that mediate  'RKKY like'  exchange.   Below gap
excitation  induces an  optical coherence between conduction and valence band
able to mediate exchange interactions  between localized
spins\cite{PC02,FP04}.   The strength of the 'optical RKKY' exchange
interaction (ORKKY) is determined  both by the intensity of the laser,
proportional to the square of  the Rabi energy $\Omega$ \cite{PC02}, and by
the detuning $\delta=E_g-\omega_L$  between the semiconductor gap and the
laser frequency. When $\delta>\Omega$, the laser-matter coupling can be
treated in perturbation theory \cite{PC02} and the strength of the coupling is
proportional to $\Omega^2/\delta^3$.

Here we study  the interaction between two spins   located in a quantum
dot  embeded in a microcavity tuned close to resonance with the quantum dot
semiconductor gap. The dot provides full three dimensional confinement for
the electrons and the holes and the cavity provides full three dimensional
confinement for the photon field. A large 
value of the electromagnetic density is obtained due to 
photon confinement  in a microcavity  without the use of 
ultra-short laser pulses \cite{nano-cavity}. The exchange between the
intermediate fermions and the localized spins is also enhanced if the former
are confined \cite{FB04}.   We find that confinement of both the light and the
intermediate fermions  yields an enhancement of the ORKKY interaction  so big
that  {\em a single photon can induce an   indirect optical exchange
interaction} between  two localized  spins at temperatures of 1 Kelvin.
Confinement also permits  the {\em exact}
diagonalization of the Hamiltonian, considering all the localized spins, the
intermediate fermions and the electromagnetic field {\em fully quantum
mechanically}.  The exact solution of the problem provides  
an unified treatment of carrier mediated and optical IEI.

   
In the following, we consider a micrometer size cylindrical cavity
\cite{nano-cavity}, made of CdTe with inclusions of (Cd,Mn)Te quantum dots
\cite{DMS-dots}. 
We consider small parallelepiped dots \cite{Barenco-dot} that confine conduction
band electrons  (creation operator $c^{\dagger}$) and valence band holes
(creation operator $d^{\dagger}$) 
with intra-band level spacing 
larger than all
the other intra-band energy scales in the problem, so that we only keep the
lowest orbital level in each band, $\epsilon_c$ and $ \epsilon_v$. These levels
have a twofold spin degeneracy. Their orbital wave functions are 
$\psi_e(\vec{r})$) and  $\psi_h(\vec{r})$ respectively.

The electric field of the lowest cavity mode lies mainly in the plane perpendicular to
the axis of the cylinder. In consequence,  there are two degenerate cavity
modes, associated to the two possible polarization states in that plane.  Their
energy  $\hbar\omega_0$  is close to the quantum dot band gap,
$E_g$.   We choose  the circularly polarized cavity modes as a basis and, after
canonical quantization, the corresponding photon creation operator is denoted
by  $b^{\dagger}_{\lambda}$, where $\lambda=L,R$. The most general Hamiltonian
we consider can be splitted in 4 terms, ${\cal H}={\cal H}_0+{\cal H}_{g}+{\cal
H}_{J}+{\cal H}_{U}$. The first reads
\begin{eqnarray}
{\cal H}_0=\sum_{\lambda} \hbar \omega b_{\lambda}^{\dagger} b_{\lambda}
+\sum_{\sigma}\left [ \epsilon_v d_{\sigma}^{\dagger}d_{\sigma}
+\epsilon_c c_{\sigma}^{\dagger}c_{\sigma}\right]
\end{eqnarray}
and describes the Hamiltonian for decoupled cavity photons and
quantum dot fermions. The light-matter coupling only  has 
 non diagonal terms in the band index:
\begin{eqnarray}
{\cal H}_{g}=
\sum_{\sigma_e,\sigma_h,\lambda}
G^{\lambda}_{\sigma_e,\sigma_h}
\left(b_{\lambda}^{\dagger}+b_{\lambda}\right)
\left[
c_{\sigma_e}^{\dagger}d^{\dagger}_{\sigma_h}+d_{\sigma_h}c_{\sigma_e}\right]
\end{eqnarray}
If we assume that the hole is purely heavy,  we obtain  the standard spin
selective coupling \cite{PC02,FP04,OO} $
G^{\pm}_{\sigma_e,\sigma_h}=\frac{g}{2}\left(\delta_{\sigma_e,\sigma_h} \pm
\hat{z}\cdot\vec{\tau}_{\sigma_e,\sigma_h}\right)$  where $g$ is the Rabi
energy and $\vec{\tau}$ are the Pauli matrices. This coupling breaks spin
rotational invariance and privileges the axis of the cavity,  $\hat{z}$.  The
value of $g$ depends on the amplitude of the cavity mode in the location of the
dot and plays the same role than the Rabi energy $\Omega$ in the case of a
photoexcited semiconductor \cite{PC02,FP04}. 
 The Hamiltonian ${\cal H}_0+{\cal H}_g$  is identical
to two independent Jaynes-Cummings models, one for each cavity mode. A key
quantity that governs this system is the detuning $\delta\equiv E_g 
-\hbar\omega$. From the 4 possible electron-hole pairs,  two 'bright'
$|X_B\rangle$ pairs are coupled to the  cavity modes ($|CM\rangle$)and
 two 'dark' 
$|X_D\rangle$ pairs are  decoupled from the rest.  The spectrum of ${\cal
H}_0+{\cal H}_{g}$ is shown in fig. 1, for different values of $\delta$
 in the manifold of ${\cal N}=1$ 
excitation (see below). The levels have a trivial $(2S+1)^2$
 degeneracy of the
uncoupled  local spins $S=5/2$. 
The  ground state manifold is mainly photonic in the 
$ \delta>0$  case (1a), mainly excitonic ($|X_B\rangle$)  in the $ \delta<0$
case (1c) and it is a compensated mixture in the $\delta=0$ case (1c) . The 
exchange interaction between the fermions and  the spin $S=5/2$ of the Mn 
impurities reads: 
 \begin{equation}
{\cal H}_{J}= \sum_{I,f}J_f \vec{S}_I\cdot\vec{S}_f\left(\vec{x}_I\right)
\end{equation}
where $\vec{S}_f\left(\vec{x}_I\right)$ stands for local spin density of the
$f=e,h$ fermion and $\vec{S}_I$ is the  Mn spin located in $\vec{x}_I$.
 The electron spin density reads:
$\vec{S}_e\left(\vec{r}_I\right)=\frac{1}{2}|\psi_e(\vec{r}_I)|^2 
c^{\dagger}_{\sigma}c_{\sigma'}\vec{\tau}_{\sigma,\sigma'}$
and analogously for the holes. The strength of the
interaction between the quantum dot fermion and the magnetic impurity depends
both on the exchange constant of the material $J_f$ and on the localization
degree of the carrier determined by the confinement of the dot,
 $|\psi_f(\vec{r}_I)|^2$. 
We consider a hard wall quantum dot, with lateral dimensions $L\simeq 10$ nm
and total volume $\Omega_{QD}\simeq 1200$  nm$^3$. The upper limit for the exchange
interaction between a conduction (valence) band  electron (hole) and a Mn spin
is $j^{max}_e=8\times \frac{J_e}{\Omega_{QD}}=-0.1$meV 
($j^{max}_h=8\times \frac{J_h}{\Omega_{QD}}=+0.5$meV ). The exchange
coupling can be recasted as 
$\frac{1}{2}\eta(\vec{x}_I) j^{max}_f\vec{\tau}_f\cdot\vec{S}_I$. For a given
dot,  $0<\eta<1$. Larger exchange interaction ($\eta>1$) can be obtained in
smaller dots.
  
The last term in the Hamiltonian
describes the intra-band $U_1$ and inter-band $U_2$ Coulomb interactions
between the  quantum dot electrons\cite{note}:
\begin{equation}
{\cal H}_{U}= U_1 \left( n_{c,\downarrow}
n_{c,\uparrow} +  \overline{n}_{d,\downarrow} \overline{n}_{d,\uparrow}\right)+
 U_2 \sum_{\sigma,\sigma'}n_{c,{\sigma}}\overline{n}_{d,{\sigma'}}, 
 \label{coulomb}
\end{equation}
\noindent where $n_{c,{\sigma}}= c_{\sigma}^{\dagger}c_{\sigma}$ and
$ \overline{n}_{d,{\sigma}}=1-d_{\sigma}^{\dagger}d_{\sigma}$. When the interaction is
included, we have $E_G=\epsilon_c+\epsilon_v-(U_1-U_2)$.

The band gap in (Cd,Mn)Te  is the largest
energy scale in the problem. We have verified numerically that two consequences
follow. First, the effect of the terms that do not conserve the number of
excitons plus photons in ${\cal H}_g$ is fully negligible and they can safely
be removed from the Hamiltonian. This is known as the rotating wave
approximation  and permits to  decompose the Hilbert space in uncoupled
and numerically tractable subspaces with ${\cal N}$ excitations. 
Second, 
the ground state of the problem, which lies in the ${\cal N}=0$ sub-space,  has a
extremely small Mn-Mn coupling due to accross-gap Bloembergen-Rowland coupling.
Here we consider the Mn-Mn coupling, in presence of 
${\cal N}>0$
polaritons and we solve {\em exactly} the Schrodinger equation ${\cal
H}|\Phi^{\cal N}_i\rangle= E^{\cal N}_i |\Phi^{\cal N}_n\rangle$.

\begin{figure}
\includegraphics[width=3in,height=3in]{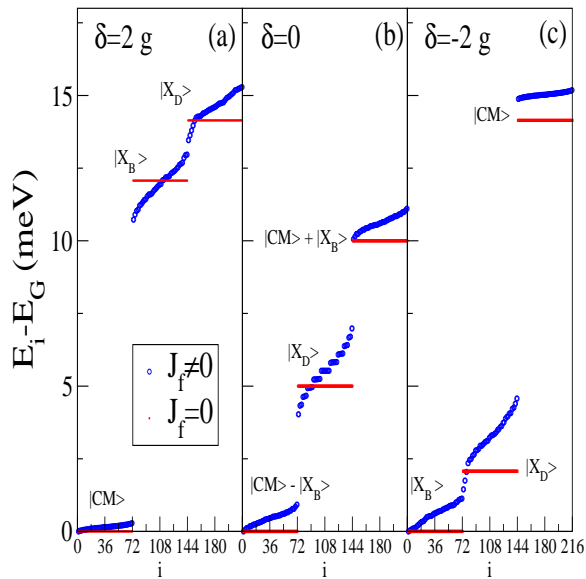}
\caption{Spectrum of ${\cal H}$ for $g=5$ meV and different values of $\delta$,
 $S$, with (symbols) and without (points) exchange. }
\end{figure}

{\em Results for ${\cal N}=1$}. In this manifold  the effect of 
  ${\cal H}_U$ is trivial. For simplicity, we take $U_1=U_2=0$
  in this case.  
Figure 1  (symbols)  shows the excitation spectrum $E_i-E_G$, where $E_G$
is the ground state of the ${\cal N}=1$ manifold and $i$ labels the excited
states, for $g=5$ meV and 3 different detunings, $\delta=-2\times g$ (fig 1a),
 $\delta=0$ (fig 1b) and $\delta=2 g$ (fig 1c). 
The flat lines correspond to the energy spectrum without exchange coupling.
The dispersion of the energy levels, compared with the
case without exchange, indicates the strength of the indirect exchange
interaction and, as seen in figure 1, is proportional to 
the excitonic content of the levels. 
 In order to quantify how the impurities are correlated we define
a spin-spin correlation:
\begin{equation}
\langle\vec{S}_1\cdot\vec{S}_2\rangle_{\cal N}=
\frac{1}{Z_{\cal N}}\sum_{i}
\langle\Phi^{\cal N}_i|\vec{S}_1\cdot\vec{S}_2|\Phi^{\cal N}_i\rangle
{\rm
e}^{-E^{\cal N}_i/kT}
\label{corr}
\end{equation}
\noindent  
where $Z_{\cal N}$ is the partition function and  the sum runs over the 
eigenstates $\Psi_i$ of the Hamiltonian with energies $E_i$ in the $ {\cal N}$
manifold. This correlation function corresponds to a density matrix in which
the degrees of freedom are in equilibrium inside the manifold with ${\cal N}$
polaritons.  
In figure 2a we plot $\langle\vec{S}_1\cdot\vec{S}_2\rangle_{{\cal N}=1}$
 as a function of $\delta$ and $g$
at $k_b T=1$K and $U_1=U_2=0$.
Results with finite values of $U_i$ can be 
  obtained exactly    by
  replacing $\delta(U_1=U_2=0)$ by $\delta(U_1=U_2=0)-(U_2-U_1)$.
 The phase diagram has 3 different regions. 
In region I ($\delta>g>0$),  the polariton is mainly photonic but, because of
the light-matter coupling $g$, it has a small excitonic component that correlates
the spins. In this region the spin correlation vanishes identically if $g=0$.
This is the optical exchange interaction region. In region (II) 
($|\delta|<g$) the polaritons
have a large content of both exciton and photon that is properly
captured by  our non-perturbative approach.
Region (III) ($\delta>g>0$) the polariton is mainly excitonic and the spin
correlation comes from  carrier mediated
exchange interaction region. Since the weight of excitonic part in the wave
function is larger than in regions (I) and (II) the interactions are
consequently stronger.  
The segment $g=0,\delta<0$ in fig. 2a (see also  2d) corresponds to a bare quantum
dot, totally decoupled from the cavity, and occupied by 1 exciton.
 In this case a significant ($S^2/3$)  correlation survives at 15 Kelvin (fig.
2c). Figures 2a and 2b support our claim that confinement
of both the cavity mode and the
excitons enhances optical
exchange interaction to the point that {\em a single photon} correlates 2
distant spins at $k_BT=1$K.

In figure 2b,2c and 2d  we  plot  $\langle S^1_zS^2_z\rangle_{\cal N}$ and
$\langle S^1_{\perp}S^2_{\perp}\rangle_{\cal N}$ as a function of temperature
for region (I) (fig. 2b), region (III) with finite $g$  (2c) and the
purely excitonic case ($g=0$, 2d). Figures 2b and 2c describe the same
system than  figures 1a and 1b respectively.
Only the case with $g=0$ (fig. 2d) has 
$2\times \langle S^1_zS^2_z\rangle_{\cal N}=\langle
S^1_{\perp}S^2_{\perp}\rangle_{\cal N}$.  This proves that ${\cal H}_{g}$  is
responsible for the anisotropy of the spin correlations seen in figs. 2b
and 2c.


\begin{figure}[h]
\includegraphics[width=3.in,height=3in]{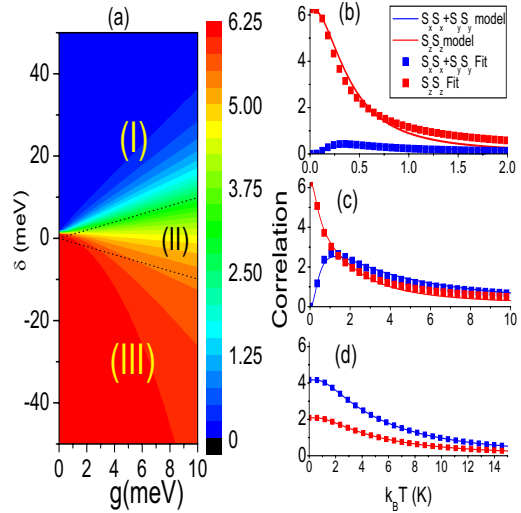}
\caption{Left: Contour plot for $\vec{S}^1\cdot\vec{S}^2$ as a function of $g$
and $\delta$ for $k_BT=1$K. Right: Correlation functions both for ${\cal H}$
 (lines) and
${\cal H}_{eff}$ (symbols) }
\end{figure}

A simple estimate of ordering temperatures in dots with many spins can be
obtained if  the low energy sector of our model is
 described with an anistropic Heisenberg  Hamiltonian  ${\cal H}_{eff}=
-J_x\left(S^1_xS^2_x+S^1_yS^2_y\right)-J_z S^z_1S^z_2 $. We find  $J_x$ and
$J_z$
by fitting the static spin correlation function of eq.({\ref{corr}) to that of
the Heisenberg model (fig. 2(b,c,d)).  Remarkably, it is always possible to
find $J_x$ and $J_z$ such that correlations functions are similar to a few
percent in a wide temperature range. The outcome of the procedure is stable
with respect to small variations of parameters in the problem.
We obtain 
$J_x=$15mK and $J_z=$0.13K for $\delta=2g=10 meV$ (fig 2b),
$J_x=$0.35K and $J_z=$0.58K for $\delta=-2g=-10 meV$ (fig 2c),
$J_x=J_z$ 0.6K for $g=0$, $\delta<0$ (fig 2d). 
Using the mean field result
$k_BT_c = \frac{S(S+1)}{3} z J$, where $z$ is the typical number of Mn coupled
to a given one, we obtain $k_B T_c \simeq z\times 2$K for the case of fig. 2d,
that corresponds to  exciton mediated coupling. 
The exciton mediated spontanenous polarization in CdTeMn
quantum dots surviving at 120 Kelvin recently reported \cite{dot-Smith}, could
be justified if we take  $z\simeq 60$ wich is not unreasonable \cite{FB04}.
 The above results are obtained for $\eta=1$. In
figure 3d we plot the correlation function for the same system of figs. 1a, 1b
and 1c as a function of $\eta$ at 1 Kelvin. For $\eta<1$ ($\eta >1$)
these curves reflect how the spin
spin correlations are degraded (enhanced) as the fermion's wave function 
is less (more) peaked on
the Mn positions. The region  $\eta >1$  corresponds to  dots smaller than the
$\eta=1$ reference.

{\em Results for $ {\cal N}>1$}. Optical exchange interaction  in bulk is
proportional to the square of the electromagnetic density.  
In the Jaynes-Cummings model the light matter coupling is renormalized like
$g_{\cal N}= {\cal N}g$ and figure 2a shows how the spin correlation is an
increasing function of $g$ in regions $(I)$ and $(III)$.
 Hence, it might seem that increasing ${\cal N}$
should increase the effective coupling.
In contrast, double occupancy of the fermion levels  in the dot 
blocks the possibility of spin exchange with the impurities and dramatically
reduces the effective interaction \cite{FB04}. The relative importance  of
these competing factors is tuned as $\delta$ changes. Interestingly, 
intra-band Coulomb repulsion reduces the double fermion occupancy and favors
the ferromagnetic exchange. Figs. 3a,3b and 3c display the spin correlation
function for $k_BT=1$ K as a function of ${\cal N}$ for the same 
cavity-dot systems of figures 1a, 1b and 1c respectively, with 
$U_1$=20 meV and $U_2=10$ meV.     
In the three cases the total correlation decreases as the number of cavity
exciation increases, reflecting that Pauli blocking overcomes the enhancement
of Rabi energy. The dominant axis of the spin correlation undergoes  
a crossover from off plane ($z$) to in plane ($xy$) 
that reflects how the change in the relative weight of the bright and dark 
excitons as the Rabi coupling increases. This crossover opens the door to 
optical tuning of the easy axis in the ferromagnetic phase. 

\begin{figure}
\includegraphics[width=2.5in,height=3.2in]{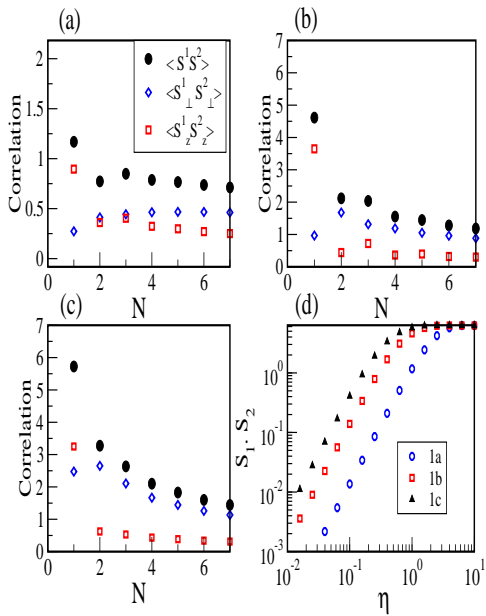}
\caption{(a), (b), (c): spin correlation as a functions of {\cal N} for the same parameters than fig. 1a,1b
and 1c respectively. (d): spin correlation function as a function of $\eta$.  
}
\end{figure}

{\em Discussion and conclusions}.
We have studied the indirect exchange interaction between two spins in a
cavity-dot system with ${\cal N}$ exciton polaritons. 
In practice the Fock cavity states could be
prepared by pumping the cavity with the adequate laser\cite{Teje}, although
some electronic injection method could be devised\cite{Injection}. 
The detection of the resulting spin correlations could be done by 
spin resolved  photoluminescence detection \cite{dot-Smith} or by pump
and probe  \cite{Merlin02}.
The exact diagonalization of our model permits to study the evolution
of the indirect exchange interaction  as  the cavity  mode energy crosses over the 
band gap, going from  the limit of 'optical RKKY' \cite{PC02,FP04} to conventional
carrier mediated interaction.  Attending
to the number of photons involved,  the cavity-dot system provides a huge
enhancement of the optical exchange interaction   compared with bulk systems
\cite{PC02,FP04}, since a single photon  yields a sizable indirect exchange
interaction that survives at 1 kelvin. This makes intense
lasers unnecessary in the implementation of the  optical exchange interaction.

\acknowledgments
We acknowledge C. Piermarocchi for fruitful discussions.
JFR acknowledges financial support from Grants,
MAT2003-08109-C02-01, Ramon y Cajal Program (MCyT,Spain),
and   UA/GRE03-15. This work  has been partly funded by FEDER funds.  
Partial financial support by the spanish MCYT (grant MAT2002-04429-C03),
the Universidad de Alicante, the brazilian agencies FAPERJ, CAPES and CNPq 
are gratefully acknowledged. 
G. Chiappe thanks UBACYT and CONICET for finantial support.


\begin{thebibliography}{99}

\bibitem{LD98}  D. Loss and D. P. Divincenzo, Phys. Rev. A{\bf 57}, 120 (1998).

\bibitem{SI99}  
A. Imamoglu {\em et al.}, Phys. Rev. Lett. {\bf83}, 4204 (1999).

\bibitem{Ka98}
B. E. Kane, Nature {\bf 393}, 133 (1998)

\bibitem{WA01} 
S.A. Wolf {\em et al.}, Science {\bf 294}, 1488 (2001). 

\bibitem{AL02}  
D.D. Awschalom, D. Loss, and N. Samarth (editors),
{\it Semiconductor Spintronics and Quantum Computation}  
(Springer, New York, 2002).

\bibitem{PC02}
C. Piermarocchi {\em et al.} Phys. Rev. Lett. {\bf 89}, 167402 (2002).

\bibitem{Ca03} T. Calarco {\em et al.}, Phys. Rev. A{\bf 68}, 012310 (2003)

\bibitem{Merlin02} J. Bao, {\em et al.} , Nature Mater. {\bf 2}, 175 (2003).
J. Bao {\em et al.} , cond-mat/0406672 

\bibitem{FB04} J. Fern\'andez-Rossier and L. Brey, {\rm cond-mat/0402140},
accepted in  Phys. Rev. Lett.

\bibitem{FP04}
J. Fern\'andez-Rossier {\em et al.},  (condmat/0312445).
 Accepted in Phys. Rev. Lett

\bibitem{Nature} H. Ohno {\em et al.}, Nature, {\bf 408}, 944 (2000).
D. Chiba {\em et al.}, Science, {\bf 301}, 943 (2003)

\bibitem{BK02} 
H. Boukari {\em et al.},  Phys. Rev. Lett. {\bf 88}, 207204
(2002).

\bibitem{dot-Smith} 
S. Mackowski {\em et al.} Appl. Phys. Lett. {\bf 84}, 3337 (2004) 



\bibitem{Ba95} A. Barenco {\em et al.}, Phys. Rev. A{\bf 52}, 3457 (1995)

\bibitem{Di00} D. P. DiVincenzo {\em et al.}, Nature {\bf408}, 339 (2000)

\bibitem{nano-cavity} 
G. S. Solomon, M. Pelton and Y. Yamamoto, Phys. Rev. Lett. {\bf 86}, 3903
(2001).
\bibitem{DMS-dots} 
A. A. Maksimov {\em et al.}, Phys. Rev.  B {\bf 62}, R7767 (2000). 
G. Bacher {\em et al.} Phys. Rev. Lett. {\bf 89}, 127201 (2002).

\bibitem{Barenco-dot} A. Barenco and M. A. Dupertuis, Phys. Rev. B{\bf 52}, 2766
(1995)


\bibitem{OO} {\em Optical Orientation}, edited by F. Meier and B. P.
Zakharchenya (North Holland, New York, 1984) 



\bibitem{note} Eq. (\ref{coulomb}) differs from the standard Hamiltonian
\cite{Barenco-dot} by a
trivial one body shift $V=U_1-U_2\sum_{\sigma}n_{c,\sigma}+n_{d,\sigma}$   
\bibitem{Teje} J. I. Perea, D. Porras and C. Tejedor, cond-mat0310570

\bibitem{Injection} A. Zrenner {\em et al.}, Nature {\bf 418}, 612 (2002) 

\end{thebibliography}
\end{document}